**Exploring Strategies for Personalized Radiation Therapy: Part *III* – Identifying genetic determinants for Radiation Response with Meta-Learning**

Hao Peng, Yuanyuan Zhang, Steve Jiang, Robert Timmerman and John Minna

Medical Artificial Intelligence and Automation Laboratory, University of Texas Southwestern Medical Center, Dallas, TX, USA

Department of Radiation Oncology, University of Texas Southwestern Medical Center, Dallas, TX, USA

Nancy B. and Jake L. Hamon Center for Therapeutic Oncology Research, University of Texas Southwestern Medical Center, Dallas, TX, USA

**Corresponding Author:**
Hao Peng, PhD, Email: Hao.Peng@UTSouthwestern.edu

**Running title:** Personalized Radiation Therapy

The authors have no conflicts to disclose.

Data sharing statement: The data that support the findings of this study are available from the corresponding authors upon reasonable request.

**Abstract**

Radiation response in cancer is shaped by complex, patient-specific biology, yet current treatment strategies often rely on uniform dose prescriptions without accounting for tumor heterogeneity. In this study, we introduce a meta-learning framework for one-shot prediction of radiosensitivity (measured by SF2) using cell line–level gene expression data. Unlike the widely used Radiosensitivity Index (RSI)—a rank-based linear model trained on a fixed 10-gene signature—our proposed meta-learned model allows the importance of each gene to vary by sample through fine-tuning. This flexibility addresses key limitations of static models like RSI, which assume uniform gene contributions across tumor types and discard expression magnitude and gene–gene interactions. Our results show that meta-learning offers robust generalization to unseen samples and performs well in tumor subgroups with high radiosensitivity variability, such as adenocarcinoma and large cell carcinoma. By learning transferable structure across tasks while preserving sample-specific adaptability, our approach enables rapid adaptation to individual samples, improving predictive accuracy across diverse tumor subtypes while uncovering context-dependent patterns of gene influence that may inform personalized therapy.

# Introduction

In precision oncology, treatment response varies widely across patients and cell lines. In the first two parts, we explored the strategies of identifying image features related to treatment outcome using class activation mapping techniques and diffusion models [1, 2]. In this part, we shift from image features to genetic biomarkers to predict radiation response, which varies for different patients. This heterogeneity reflects complex, patient-specific gene expression patterns and highlights the need for models that can personalize predictions. Each cell line is a unique biological "instance" of cancer, representing a different patient or tumor subtype. A gene that plays a crucial role in one sample might be different in another. This raises an interesting question, whether can we identify which genes are most relevant for each individual patient, rather than relying solely on group-level averages?

This is precisely the type of problem that meta-learning is designed to solve [3-6]. Meta-learning facilitates fast adaptation from limited data, making it well-suited for scenarios where each tumor should be viewed as a distinct task. In our study, we have the experimental results of radiosensitivity (survival fraction, SF2) of 73 cell lines. SF2 depends on multiple tumor intrinsic factors, such as DNA damage repair capacity, cell cycle checkpoints, hypoxia and metabolism, which is regulated genetically and epigenetically. When using a standard regression model (including neural networks), we only obtain a single representation across all cell lines and lack the ability to adapt to a new individual. This offers limited personalization and compromise predictive accuracy, given the relationship between genetic features and SF2 is extremely complex. In other words, global models reflect average effects across the entire population, identifying what works best on average. However, they may overlook genes that are especially important in specific subgroups or fail to capture nonlinear relationships and gene-gene interactions that only emerge in certain individuals.

Meta-learning offers a promising alternative by training models to *learn how to learn*. Rather than optimizing for performance on a fixed set of training samples, meta-learning frameworks (e.g., MAML, REPTILE) optimize for rapid adaptation to new tasks with only a small number of examples per task [5, 6]. When tasked to predict radiation response (SF2) by genetic profiling, each tumor can be viewed as a separate task. The model is trained across many such tasks to develop a good initialization that can quickly adapt to a new sample with minimal fine-tuning. After adaption in the interference phase, the unique relationship between gene expression and radiosensitivity can be established. More importantly, this allows us to further investigate gene relevance for each tumor through feature sensitivity or importance.

To the best of our knowledge, this is the first study to apply a meta-learning framework to predict SF2 based on gene features. Previous work has largely centered on the Radiosensitivity Index (RSI) and conventional statistical modeling [7-13]. For instance, one study modeled SF2 in 48 genetically annotated human cancer cell lines, identifying a 474-gene interaction network that included key regulators of DNA damage repair (e.g., *ATM*, *XRCC6*), oxidative stress (e.g., *PRDX1*, *TXN*), and major signaling hubs (e.g., *JUN*, *STAT1*, *RELA*, and *IRF1*) [7]. The RSI has been independently validated as a predictor of radioresponsiveness across multiple tumor types by multiple studies [8-11], as well used to introduce the concept of genomic-adjusted radiation dose (GARD) [12, 13]. In the emerging field of combined radiation and immunotherapy, the variability in radiosensitivity has also prompted the use of SF2 and RSI to better understand the interplay between treatment response and tumor immune profiles. One study analyzing 10,469 transcriptionally profiled tumors across 31 cancer types revealed significant heterogeneity in both radiosensitivity and the immune microenvironment [17]. Tumors predicted to be more radiosensitive exhibited enriched interferon signaling, higher infiltration of immune cells such as $CD8^+$ T cells, activated natural killer cells, and M1 macrophages, as well as diverse expression patterns of immunoregulatory genes. Another study applied a regression model based on RSI (derived from 10 genes) to assess whether genomic indicators of radiosensitivity correlate with immune activity [18].

In this exploratory study, we take a first step toward modeling the relationship between transcriptomics and radiation response. We begin with conventional approaches—standard least squares regression and basic neural networks without meta-learning—as baselines. These methods assume a global relationship across all samples and do not account for biological heterogeneity. In contrast, meta-learning offers a promising path forward by treating each lung cancer line as a distinct task, allowing the model to adapt to sample-specific patterns. Training a separate model for each of the 73 cell lines is impractical due to limited data, but meta-learning algorithms such as MAML and REPTILE address this by learning a shared initialization that can be quickly fine-tuned with just a few data points. This initialization captures generalizable patterns across cell lines—such as common gene expression signatures associated with radiosensitivity—and enables personalized adaptation for new or rare samples.

**Results**

Among the 73 cell lines, SF2 values were compared across four histological subtypes of Non-Small Cell Lung Cancer (NSCLC): adenocarcinoma, squamous cell carcinoma, large cell carcinoma, and adenosquamous carcinoma, ordered from low to high (**Fig. 1**). Each subtype exhibited variability in radiosensitivity, with particularly wide SF2 distributions observed in the adenocarcinoma and large cell carcinoma groups. The top 30 genes derived from mRNA expression experiments were selected through feature selection with Lasso algorithm [19].

In a toy example of sine curve regression, the meta-learning algorithm FOMAML demonstrates faster adaptation compared to standard gradient descent from a random initialization (**Figs. 2A–2D**). Because this example uses an analytical gradient of the sine function, the model converges more easily. In contrast, with REPTILE and a neural network-based setup (**Fig. 2E**), we observe that although fine-tuning improves performance with different training steps (e.g., 100, 200, 500, 1000), the performance is still inferior compared to the case of using analytical gradient. This difference arises because, unlike the simple sine fitting case, the neural network must approximate sinusoidal behavior using many parameters—without an explicit representation of amplitude and phase. Despite this, Fig 2 shows that the meta-learned initialization is indeed able to capture a common structure across tasks: specifically, that all target functions are sinusoidal with varying amplitudes and phases. This serves as a "prior belief" that the underlying functions are smooth and periodic, even though their exact forms differ from task to task.

For the SF2 dataset, the results of REPTILE are shown in **Fig. 3** using a standard 65/8 train-test split. Across five different random splits, the training and testing loss curves demonstrate clear signs of overfitting—the model performs well on the training data but fails to generalize to unseen samples. Fine-tuning is not applicable in this setting, as each cell line contains only one sample. Without a meta-learning framework, the model lacks the ability to adapt its predictions. As a result, there is a noticeable discrepancy between predicted and actual SF2 values, highlighting the limitations of non-meta-learning approaches in this context.

**Figure 4** presents the baseline results using standard least squares linear regression with the top 20 selected features. The predicted versus actual SF2 values show substantial discrepancies across the full range of SF2 (0 to 1), indicating poor regression performance. The model achieves a mean absolute error (MAE) of 0.07 and a maximum error of 0.24. These results suggest that a linear regression model with a fixed set of gene features fails to accurately predict SF2, underscoring the need for personalized modeling tailored to each individual cell line.

**Figure 5** summarizes the REPTILE modeling results for the SF2 dataset. **Figure 5A** shows the training loss curve for 65 cell lines in the testing dataset, with convergence occurring after approximately 3000 iterations. **Figure 5B** illustrates predictions on the test set (8 cell lines) across three different random splits, comparing performance before and after fine-tuning with 10 gradient updates. The results demonstrate that meta-updates enable "personalization", shifting predictions closer to the ground truth. **Figure 5C** illustrate the effect of fine-tuning with 5, 10, and 20 steps in the meta-updates for one split (8 cell liens). The results suggest no significant overfitting. Note that light fine-tuning carries minimal risk, while excessive fine-tuning may cause the model to overfit the single test sample—memorizing the output without learning meaningful relationships between input features and SF2. In this study, overfitting is mitigated through the use of small learning rates and the shallow neural network architecture. In addition, given that the SF2 regression task is relatively a smooth line, overfitting is not expected to be pronounced.

**Figure 6** presents predicted versus actual SF2 values generated by the meta-learning model (REPTILE, 10 meta-updates) across different random data splits, with each color representing training and test samples. The model shows noticeably improved performance compared to **Fig. 4**, achieving an MAE of 0.007. Except for a few samples at the very low and very high ends of the SF2 range, the predictions closely match the true values, demonstrating significantly enhanced regression accuracy. This improvement highlights the benefit of the meta-learning approach, which allows the model to personalize its predictions by adapting gene-level weights for each task.

The "personalized" behavior of the model is illustrated in **Figure 7**, which displays the gradient of each gene for four subgroups, each ordered from low to high SF2 (top to bottom). To simplify analysis, gene expression levels are not included, allowing a clearer view of how the model internally attributes importance to each gene. A number of mechanistic interpretations are provided below. A positive gradient indicates that increasing a gene's value pushes the model to predict higher SF2, while a negative gradient suggests the opposite. Gradients near zero imply that the gene has little influence on the prediction for that particular cell line. These maps reveal that the same gene can have very different roles across cell lines—variations appear not only in the direction (sign) of the gradients but also in their magnitudes, reflecting differences in gene impact. For example, in cell line 1, the model may depend strongly on gene A, implying it plays a key role in radiosensitivity prediction. In contrast, in cell line 2, gene A may have minimal or even opposing influence, due to differences in the radiation response pathways. These observations underscore the capacity of the meta-learning model to personalize predictions based on the unique characteristics of each sample. Such maps enable direct comparison among different cell lines and answer several questions such as which features are most influential? Are certain features consistently important across many cell lines, or are some only relevant within specific subgroups?

Except for the large cell carcinoma subgroup, where gradient patterns remain relatively consistent across samples, the other three subgroups—adenocarcinoma, squamous cell carcinoma, and adenosquamous—show more complex and variable behavior. Cell lines are sorted by predicted SF2 in **Fig. 7**, and the gradients for individual features appear scattered per column—an expected outcome given the complex, nonlinear behavior of neural networks. The models inherently capture intricate feature interactions and nonlinear relationships. While a feature may exhibit a consistent overall trend across the 73 cell lines, its influence can vary significantly at the individual level—being strongly predictive in some cases, weak or negligible in others, or even exerting an opposite effect. This variation may be due to intrinsic variation of the contribution of a single gene across cell lines, or may stem from inter-feature dependencies. For instance, when the impact of gene A depends on the expression level of gene B, the relevance of gene A becomes context-dependent, leading to irregular heatmap patterns after sorting.

**Discussion**

Personalized radiation therapy aims to tailor treatment strategies to individual patient characteristics. In addition to radiomics or dosiomics features [1, 2], identifying the specific genes can further help predict therapeutic response or radioresistance in each patient (**Fig. 1**). Among patients receiving the same therapy, the underlying gene expression can differ significantly. Models that ignore this and rely only on average patterns across a cohort risk missing these individualized mechanisms. Our proposed approach moves beyond one-size-fits-all modeling, toward a more biologically faithful analysis of gene roles tied to treatment response. Moreover, the approach can also be expanded to proteomics or functional assays, which provide additional biological features [20-24]. Ultimately, the modeling outcome equips clinicians with tools to more effectively decode individualized gene expression patterns, identify meaningful biomarkers for each patient, and advance precision medicine. Meanwhile, we would like to emphasize that this study is primarily focused on developing and evaluating a novel methodological framework, rather than identifying definitive gene signatures or uncovering specific biological mechanisms. The features identified through LASSO are not guaranteed to be unique or optimal, particularly if alternative feature selection methods are used. Nonetheless, the selected subset helps reduce the dimensionality of genetic features, making the framework more practical for predicting SF2. A more in-depth biological interpretation—such as validating specific gene drivers of radiosensitivity—is beyond the scope of this work and would require further experimental validation.

The results in this study clearly reveal distinct advantages of meta-learning in terms of personalization and adaptation. Conventional regression and meta-learning represent two fundamentally different paradigms for model training and generalization [3-6]. There are over 1,000 cell lines, each representing a unique biological "instances" of cancer or tumor subtype. Meta-learning captures not only these shared patterns but also how they vary across cell lines. Certain gene expression patterns may commonly signal resistance or sensitivity to radiation. When a new cell line is introduced, the model requires only very few labeled examples for rapid fine-tuning — not from scratch, but from a meta-learned initialization. This is the strength of meta-learning: it enables fast adaptation while avoiding overfitting, even with extremely limited data. In contrast, standard supervised learning struggles with such small sample sizes (**Figs. 3 and 4**). In the current study, since only one label is used, our focus is on investigating the overall regression performance and individual sensitivity of gene features (**Fig. 6**). If our framework were extended to predict SF2, fine-tuning would be necessary at the subgroup level. In that scenario, we could treat subgroups of cell lines from the same tumor type — or across different tumor types — as a single task. By doing so, meta-updates can be better tailored to those subgroups, allowing SF2 predictions for new samples within each group.

We use gradients as an intuitive method to estimate the sensitivity (or importance) of each gene in a model's specific prediction per cell line. This approach offers valuable insight into how the model differentially weighs genetic features to arrive at a decision. The resulting gradient summarizes this process by highlighting which inputs the model is most sensitive to (**Fig. 7**). In neural networks, inputs are transformed through multiple nonlinear layers (2 layers in our study). As a result, the relationship between an input and the final output is not linear. Looking only at the first-layer weights does not capture the full impact of that feature on the prediction. Compared to alternatives like input × gradient, using the raw gradient alone helps isolate model sensitivity from variations in gene expression levels, which may be noisy. In other words, when the gradient and input × gradient diverge, the gradient alone is often more informative for interpreting model behavior. One subtle point for interpreting the patterns in **Fig. 7** is that gradients quantify the local sensitivity of the model's output to small perturbations in individual genes. However, capturing how one gene's total influence depends on another requires second-order information (e.g. the Hessian matrix). Although such dependencies are learned implicitly by deep networks, they are not directly exposed by first-order gradient analysis and remain inaccessible in the current framework.

While a direct comparison with the widely used RSI [7–11] will be the subject of future work, we highlight several key differences and potential advantages of our approach. First, RSI is based on a rank-based linear regression over 10 genes in 48 cancer cell lines and applies fixed gene weights across all samples and tumor types. This assumes gene contributions to radiosensitivity are constant across biological contexts. In contrast, our meta-learning framework allows sample-specific adaptation, enabling gene influence to vary through gradient-based fine-tuning. Second, RSI transforms gene expression into ranks within each sample, discarding expression magnitudes and potentially losing important biological signals. Our model operates directly on raw gene expression values and captures both magnitude and gene–gene interaction effects through its neural network structure. Third, RSI is static: once the model is trained, it cannot be adapted or fine-tuned for new or rare tumor types. Our approach enables rapid personalization for unseen cell lines through few-shot learning, which is especially valuable in high-heterogeneity settings, such as adenocarcinoma or large cell carcinoma (**Fig. 1**), where RSI may underperform. Finally, RSI provides no mechanistic insight at the individual level. In contrast, our method offers interpretable gradients that act as proxies for gene influence in each prediction. These gradients reveal how the same gene can play different roles across samples—positively, negatively, or negligibly—highlighting the heterogeneous nature of radiosensitivity mechanisms.

Several challenges remain. Radiosensitivity is likely driven by diverse, context-specific mechanisms—such as deficiencies in DNA repair, heightened oxidative stress, or activation of pro-survival signaling—which can vary widely across cell lines and tumor types. When a radiosensitizing gene is lost or suppressed, tumor cells may compensate through alternative pathways. Even among genetically similar cell lines, differences in transcription factor activity can lead to markedly different radiosensitivity profiles. As such, a gene may contribute to radiation response in one context but have little or no relevance in another. All these confound biological interpretations. Furthermore, establishing causality, rather than correlation, require further experimental validation—for instance, by comparing gene expression in radiotherapy-treated versus untreated samples or through gene knockout studies to assess radiation response. While our model does not resolve these biological complexities, it may serve as a flexible and individualized computational approach for predicting radiosensitivity and formulating meaningful hypotheses.

**Method**

**Data processing.** Seventy-three patient tumor–derived cell lines with detailed clinical and molecular annotations were analyzed. Radiation sensitivity was assessed using the survival fraction after 2 Gy (SF2), determined by clonogenic survival assays. Transcriptomic profiling identified a total of 46,911 gene features, of which 27,128 protein-coding genes expressed in at least 50% of the cell lines selected for modeling. The LASSO (Least Absolute Shrinkage and Selection Operator algorithms) was applied to select top 30 gene sets for SF2 regression. Owing to the high dimensionality of the gene features, most features should be eliminated to avoid overfitting. A mixture of feature selection strategies was employed. First, the features with low variance (<0.001) were excluded. Second, Pearson linear correlation analysis was performed to discard any redundant feature with high absolute correlation coefficient to any of its previous features. Third, to mitigate the potential overfitting and bias due to the small data size, a 5-fold cross-validation procedure with 10 repeats was executed for further feature selection. During each repeat of 5-fold cross-validation, the 71 cell lines were randomly shuffled and split into five parts. In each iteration, one of the parts was reserved as the test set and the remaining four parts formed the training set. The training set in each iteration was then imported into the LASSO algorithm. A fixed number of 30 non-zero coefficients were determined by tuning the regularization parameter, leading to a reservation of 30 corresponding feature candidates. Features were ultimately selected and ranked based on their importance scores averaged through 50 iterations, based on the absolute coefficients. The pair-wise correlation heat map and Variance inflation factor (VIF) were used to examine the correlation and multicollinearity issues. All analyses were conducted using MATLAB R2023b and Python v3.9.18.

**Least square fitting.** For the regression analysis, we began with a standard linear model using the top 20 features to mitigate potential overfitting, incorporating all 73 samples. Model performance was evaluated using mean absolute error (MAE), which served as a baseline for comparison with subsequent meta-learning results.

**REPTILE.** We selected Reptile as our meta-learning algorithm because it learns a good initialization for the model parameters θ, enabling fast adaptation to new tasks with just a few gradient steps. Importantly, Reptile approximates a second-order meta-gradient using only first-order information—avoiding the need to explicitly compute the Hessian matrix. In the section below, we describe how parameter updates accumulate over five inner-loop gradient steps. Suppose we take multiple gradient steps with learning rate α and let $g_i$ denote the gradient at step i. For an inner loop with five update steps, the process unfolds as follows:

$$(\theta_{k+1} - \theta_k) = -\alpha g_k \qquad \theta_5 = \theta_0 - \alpha g_0 - \alpha g_1 - \alpha g_2 - \alpha g_3 - \alpha g_4$$

The Meta update outside of the inner loop is:

$$update := \beta(\theta_5 - \theta_0) = -\beta\alpha(g_0 + g_1 + g_2 + g_3 + g_4)$$

Each gradient update can be expanded using a first-order Taylor approximation. We assume that the Hessian H remains approximately constant—which is evaluated at the initial parameter θ₀ and does not change significantly during the update steps. This assumption is reasonable when the step size α is small and the number of steps is limited (e.g., 5–10).

$$g_k \cong g_0 + H(\theta_k - \theta_0)$$

**Step 1:**

$$(\theta_1 - \theta_0) = -\alpha g_0$$

**Step 2:**

$$g_1 \cong g_0 + H(\theta_1 - \theta_0) = g_0 - \alpha H g_0 \qquad \theta_2 = \theta_1 - \alpha g_1 = \theta_0 - \alpha g_0 - \alpha(g_0 - \alpha H g_0)$$

$$\theta_2 - \theta_0 = -2\alpha g_0 + \alpha^2 H g_0$$

**Step 3:**

$$g_2 \cong g_0 + H(\theta_2 - \theta_0) = g_0 + H(-2\alpha g_0 + \alpha^2 H g_0) \qquad \theta_3 - \theta_2 = -\alpha g_2$$

$$\theta_3 - \theta_0 = -3\alpha g_0 + 3\alpha^2 H g_0 - \alpha^3 H^2 g_0$$

As shown here, after two steps, the term $\theta_2 - \theta_0$ approximates a second-order update, incorporating both the first-order descent direction $-g_0$ and a second-order correction term $-H g_0$. This behavior resembles a quasi-Newton method, where the update direction and magnitude are adjusted based on how the gradient changes—capturing curvature effects without explicitly computing the Hessian. This implicit correction is important: it prevents the model from moving too aggressively in directions with high curvature, promoting more stable learning across tasks.

**Model training.** We employed a fully connected neural network with two hidden layers (64 and 32 units) and a sigmoid-activated output to model the regression task. Input features consisted of 20 normalized genetic features per cell line. Input features were normalized to the [0, 1] range using min-max scaling based on the full dataset statistics. The data was split into training and test subsets with a 90/10 ratio. We trained the model using the REPTILE, iteratively updating model parameters through multiple inner-loop gradient updates on a batch of tasks, followed by a meta-update step aggregating these adaptations. The training proceeds as follows: 1) model weights $\theta$ are initialized randomly; 2) inner loop: for each meta-iteration, a batch of 8 tasks are randomly selected and the model undergoes 10 gradient descent steps using stochastic gradient descent (learning rate α=0.01), minimizing mean absolute error (MAE). 3) Meta update: the difference between the adapted weights and initial weights is computed per batch and accumulated as a meta-gradient. The model weights are then updated with a step size β=0.001. This meta-training process is repeated for 3000 iterations, with the step sizes linearly annealed over time. Optimization was performed using the SGD optimizer. For evaluation, the model predictions were obtained after fine-tuning gradient via the inner-loop update (10 steps) on test samples. Training and evaluation were implemented in TensorFlow 2.x with Keras API. For comparison, we implemented a non-meta-learning approach, where the model was first trained on the training set (65 samples) and then fine-tuned on the test set (8 samples) across five different random seeds and splitting.

**Toy example (sine curve regression task).** To test the feasibility, we started with the widely used toy example in meta-learning, sine wave regression. Each task involves learning a sine wave of the form *y=Asin(x+B)*, where amplitude A and phase B vary across tasks. A standard model, when trained on many such tasks, tends to average them to get a single set of parameters which do not fit individual tasks well. In contrast, meta-learning focuses on learning an initialization that can quickly adapt to any sine wave given just a few data points. This means it can rapidly adjust amplitude and phase to fit a new task after limited exposure. We tested both first-order MAML (FOMAML) and REPTILE, one with known analytical gradients (know the sine as prior) and one with a neural network.

**Feature importance/sensitivity.** Once model training is complete, we obtain a generalizable initialization that can rapidly adapt to individual patients through fine-tuning, with a single measured SF2 as the label. To interpret model predictions at the individual level, we analyze which genes the model relies on for each specific patient. Two commonly used approaches for identifying important genes are: gradients alone and gradient × input. The gradient reflects sensitivity and answers the question: "If a gene feature is increased slightly, how much will the SF2 change?" The gradient × input method accounts for the magnitude of each gene's expression, indicating how much of that feature is present in the patient relative to other genes. In this study, we employed the gradient-only approach to simplify interpretation, as gene expression levels may be influenced by experimental noise and their relationship to model predictions are nonlinear. Gradients thus offer a more stable and direct measure of the model's reliance on individual genes.

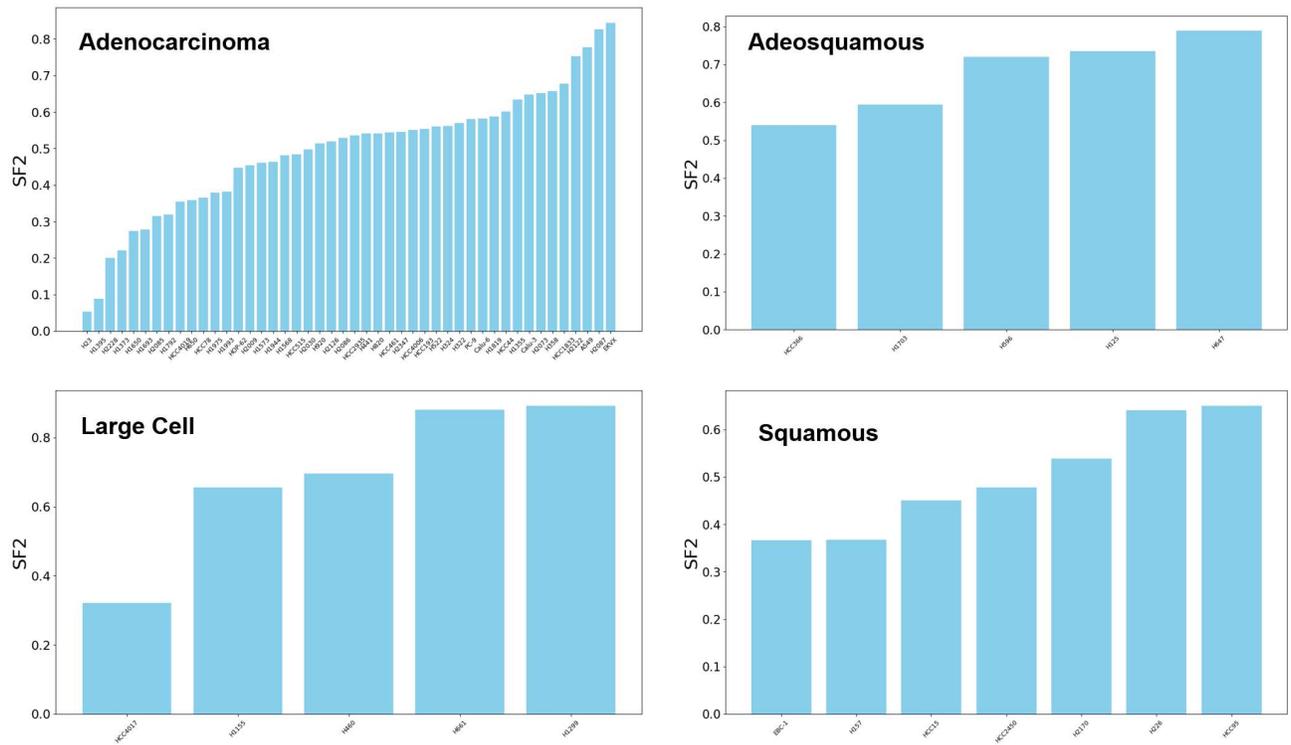

**Fig. 1.** Measured SF2 results for four tumor subgroups.

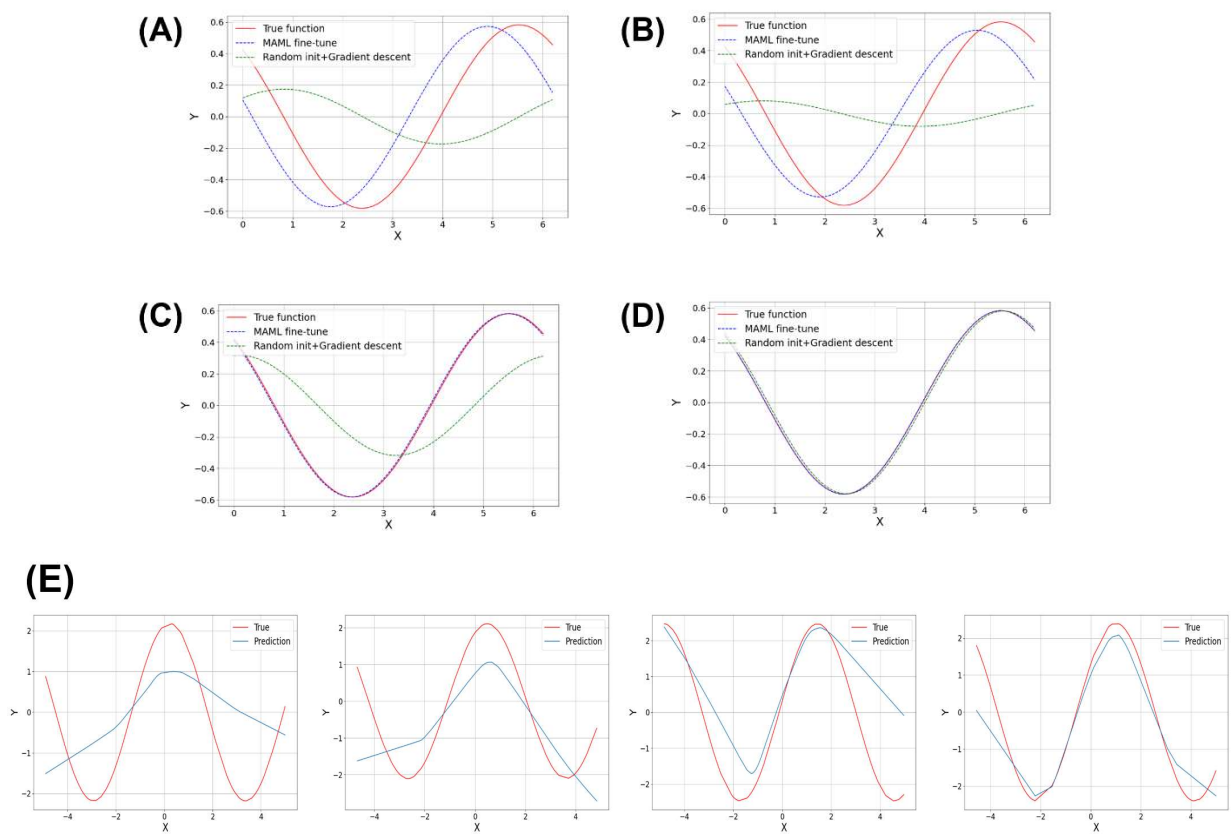

**Fig. 2.** Performance summary on the sine wave toy example comparing first-order MAML with random initialization. Each task is defined by a sine function y=Asin (x+ϕ), where A and ϕ vary across tasks (20 tasks in total, 10 support and 10 query points per task). (**A–D**) Performance using analytical gradients of sine functions with different meta-training steps (10, 100, 500, and 1000). MAML enables faster adaptation than random initialization. (**E**) Reptile results using a neural network (2-layer, 32 neurons per layer) trained on different epochs (100, 200, 500, 1000, one task per epoch, 20 data points per task), for a new unseen sample. Its performance is inferior to the results shown in A–D, primarily because the model is not explicitly trained to recover A and ϕ (i.e., without the prior knowledge that the target function is a sine wave).

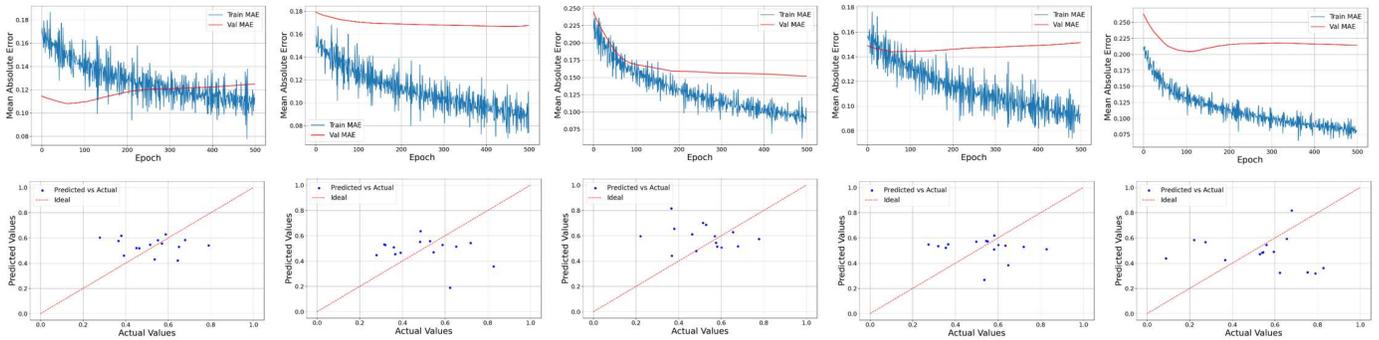

**Fig. 3.** For the SF2 dataset using a non–meta-learning approach, the model was first trained on the training set (65 samples) and then fine-tuned on the test set (8 samples) across five different random seeds and splitting. Note that since there is only one sample per cell line, fine-tuning is not feasible at the individual cell line level. (**Top**) Loss curves for both training and validation sets. Clear signs of overfitting are observed. (**Bottom**) Predicted versus actual SF2 values for the dataset, showing poor performance.

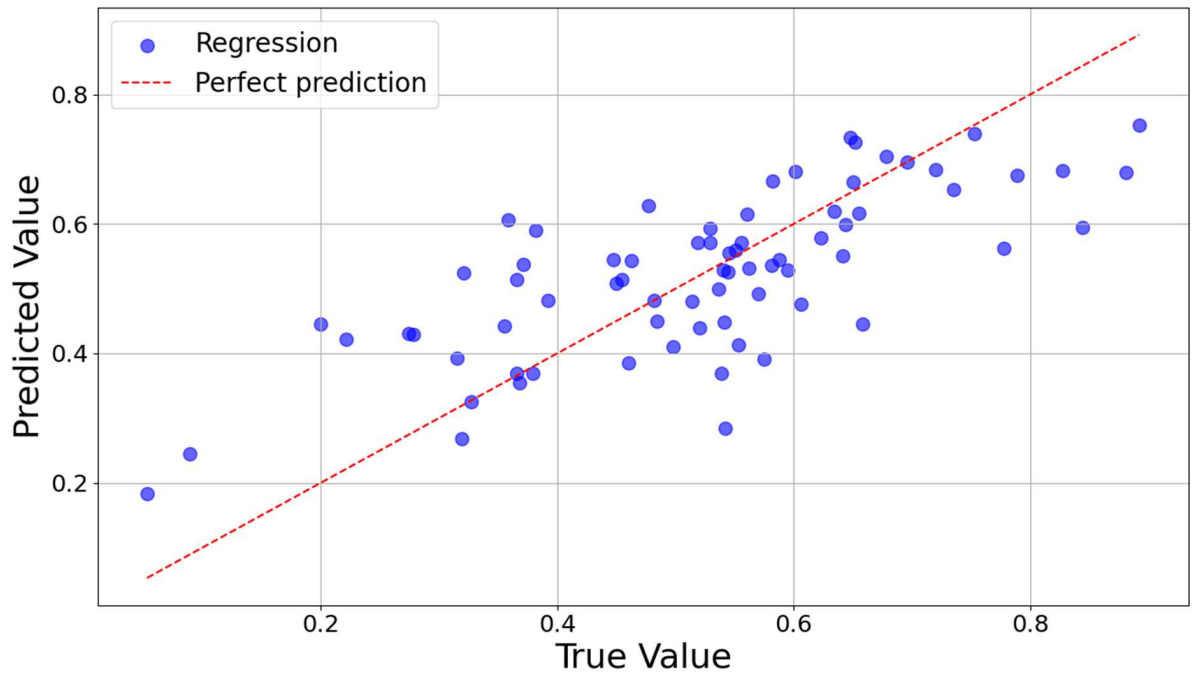

**Fig. 4.** Predicted versus actual SF2 values for the dataset using a linear regression model, achieving a mean absolute error (MAE) of 0.07 and a maximum error of 0.24.

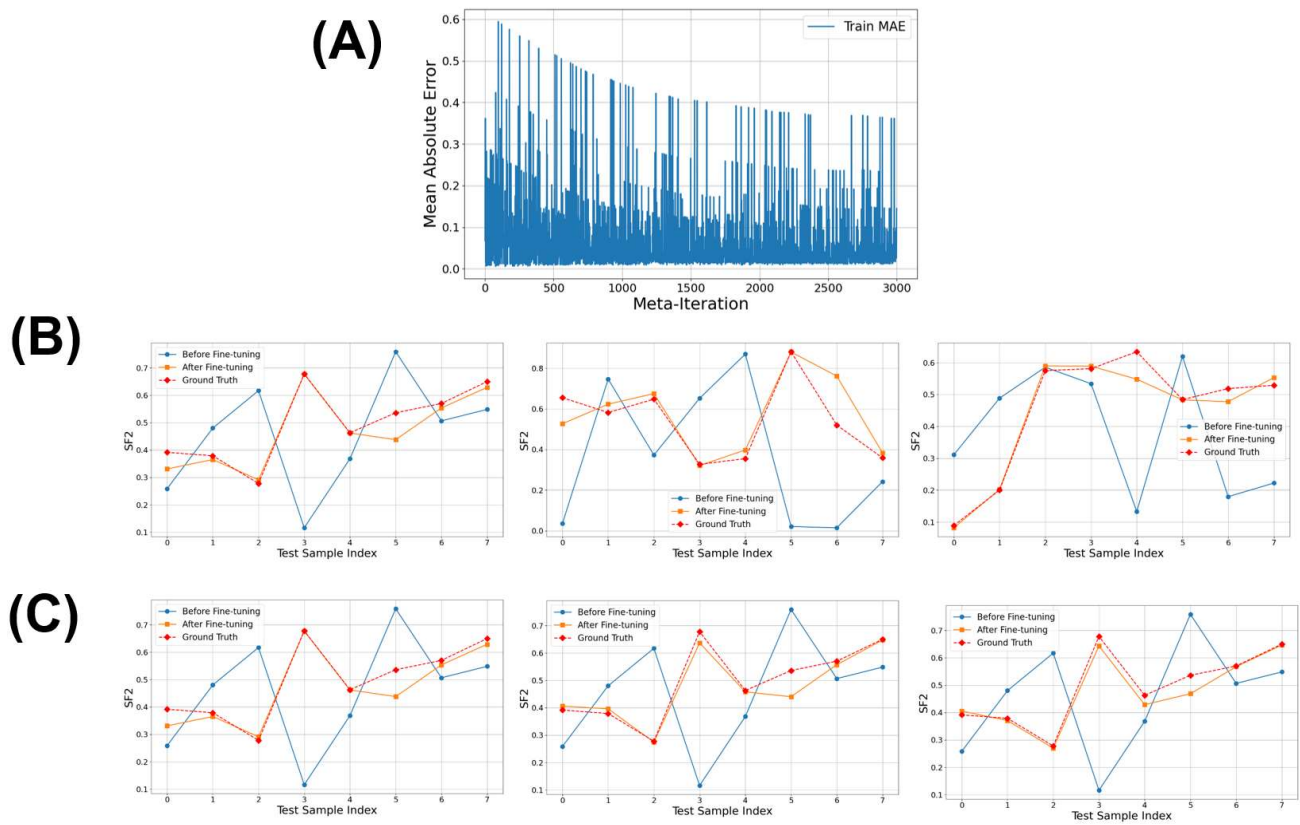

**Fig. 5. (A)** Loss curve of the Reptile model for the SF2 dataset, trained on the training set (65 samples). The loss converges after approximately 3000 iterations. **(B)** The prediction of the test set (8 cell lines) for three different random splits, showing the differences before and after fine-tuning after 10 gradient updates. **(C)** Fine-tuning results with 5, 10, and 20 inner steps suggest no overfitting.

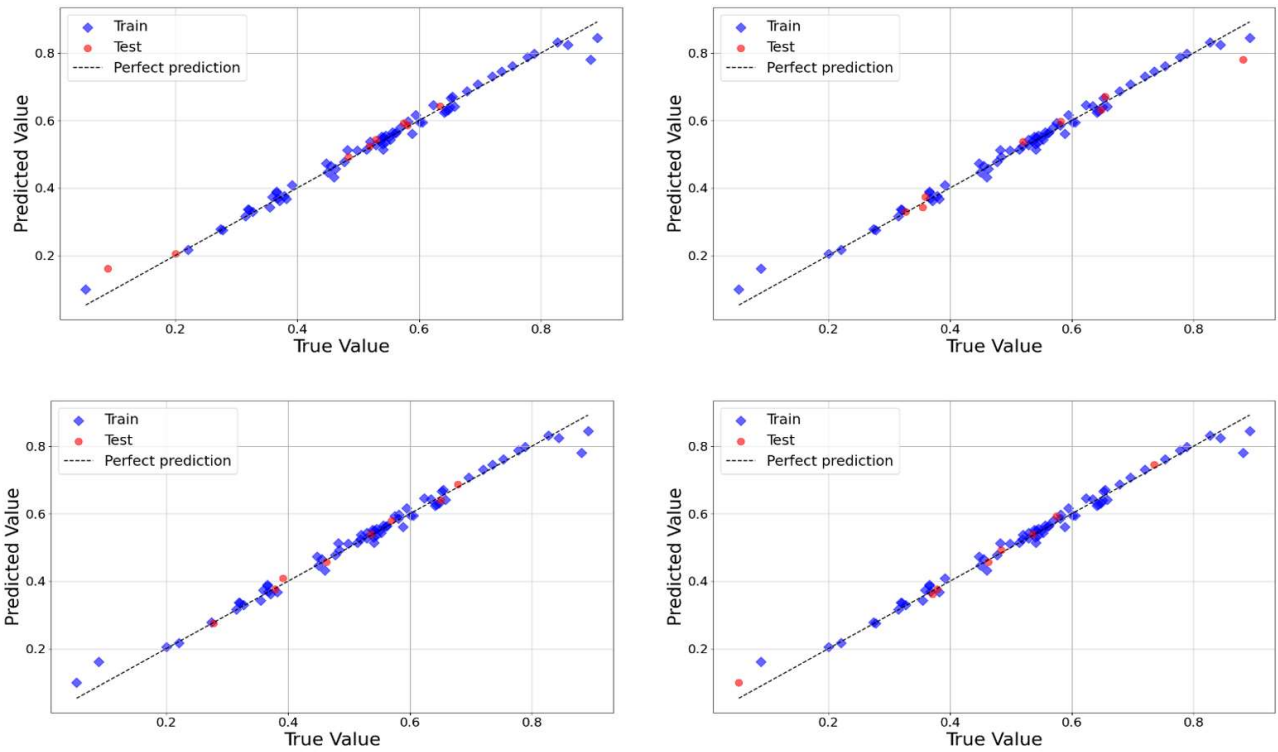

**Fig. 6.** Predicted versus actual SF2 values for four different models with different splits (different colors representing training and test samples). The model achieves noticeably better results than Fig. 4, with a mean absolute error (MAE) of 0.007.

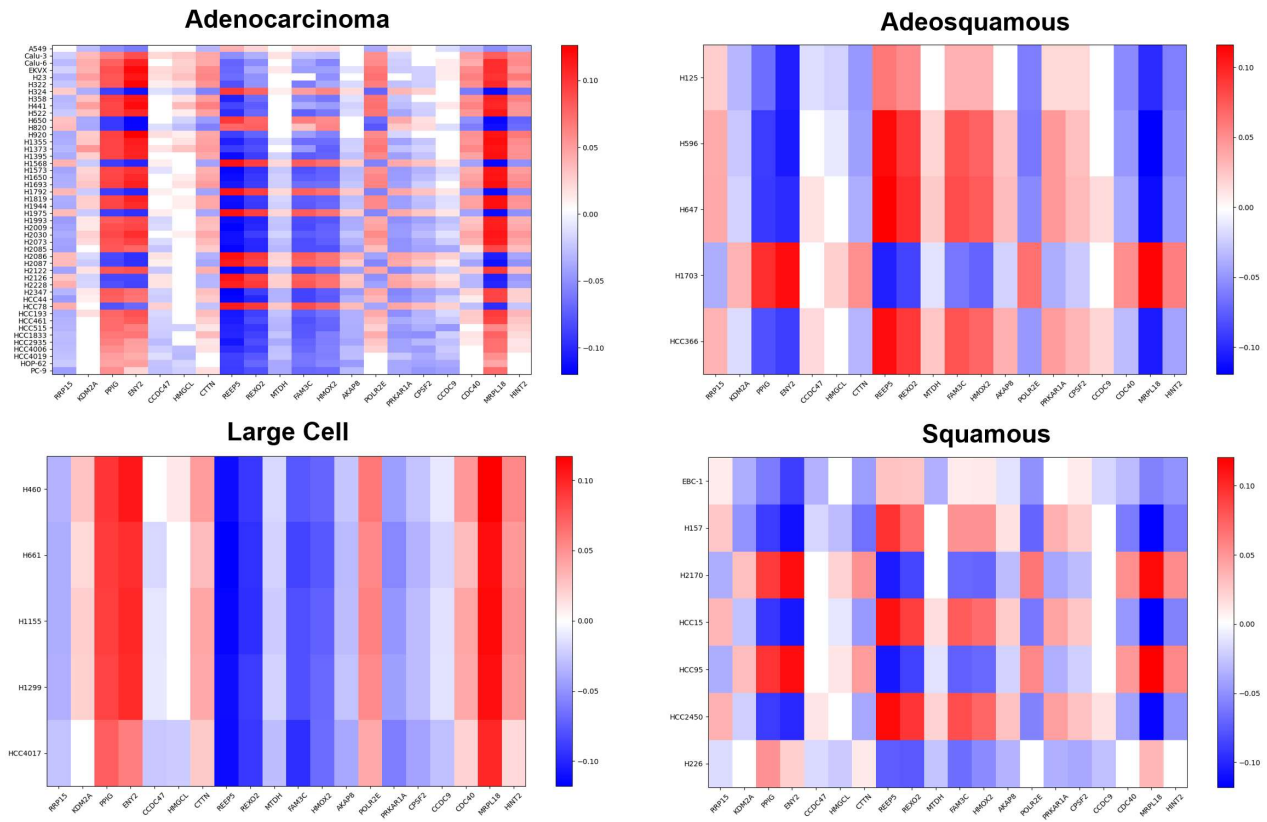

**Fig. 7.** Sensitivity analysis of the top 20 genes for four subgroups.